\documentclass{article}
\usepackage[utf8]{inputenc}
\usepackage{authblk}
\usepackage{amsmath}
\usepackage{amssymb}
\usepackage{verbatim}
\usepackage{amsthm,amsfonts,bm}
\usepackage{MnSymbol}
\usepackage{cite}
\usepackage{mathtools}
\usepackage{graphicx}
\usepackage{color}
\usepackage[colorlinks=true, allcolors=blue]{hyperref}
\usepackage{mathrsfs}
\usepackage{indentfirst}

\newcommand{\ve}{\varepsilon}
\definecolor{Blue}{rgb}{0,0,0.9}
\definecolor{Red}{rgb}{0.9,0,0}
\definecolor{Green}{rgb}{0,0.4,0}

\newcommand{\al}{\alpha}

\title{Classical gauge principle - From field theories to classical mechanics}

\author[1]{B. F. Rizzuti\footnote{brunorizzuti@ice.ufjf.br}}
\author[2]{G. F. Vasconcelos Jr.\footnote{guilherme.vasconcelos@ime.usp.br}}
\affil[1]{Departamento de Física, Universidade Federal de Juiz de Fora, MG, Brazil}
\affil[2]{Instituto de Matemática e Estatística, Universidade de São Paulo, SP, Brazil}
\date{}                     
\begin{document}
	
	\maketitle
	
	\begin{abstract}
		In this paper we discuss how the gauge principle can be applied to classical-mechanics models with finite degrees of freedom. The local invariance of a model is understood as its invariance under the action of a matrix Lie group of transformations parameterized by arbitrary functions. It is formally presented how this property can be introduced in such systems, followed by modern applications. Furthermore, Lagrangians describing classical-mechanics systems with local invariance are separated in equivalence classes according to their local structures.
	\end{abstract}
	
	\section{Introduction}\label{sec.intro}
	The importance of local invariance in field theoretical models is unquestionable. To illustrate that, we could use, for instance, the gauge principle. It states that a Lagrangian field model invariant under the action of a Lie group with global parameters can be reformulated if we raise the parameters to local ones. That is the basic ingredient R. Utiyama \cite{utiyama.PhysRev.101.1597} has used to generalize the seminal previous paper by Yang and Mills \cite{yang.mills.PhysRev.96.191}. The introduction of local parameters into the game implies the insertion of a covariant derivative, as well as a connection, in order to keep the desired invariance. In turn, it leads the initial model to interact with a gauge field. This is the very structure one finds in the unified Standard Model \cite{baez_algebra_2010}. In this case, the gauge group is given by the unitary product $SU(3) \times SU(2) \times U(1)$, setting together fermions, electroweak boson, gluon and Higgs fields.
	
	The introduction of the gauge principle dates back to 1918, when Hermann Weyl presented his ideas of building a unified field theory, which gave rise to the notion of gauge invariance \cite{Weyl1918}. Although his ideas were mathematically solid, they did not sound so physically appealing. In effect, his proposal predicted that if one transported two identical measuring rods, from points $p_1$ to $p_2$, along different paths which cross regions with distinct electromagnetic fields, generally they would no longer have the same length if measured on the point of arrival. Nonetheless, with the ascendance of quantum mechanics, his ideas were no longer physically implausible. They were actually suitable to link electromagnetism and matter fields and thus played a central role in the development of such areas. It is also worth mentioning that his seminal work still reverberates and provides the necessary background for attempts to relate gravity and electromagnetism \cite{Lindgren2021}.
	
	Weyl's proposal was based on the following. In accordance with the theory of General Relativity, spacetime is described by a $4-$dimensional manifold $M$ endowed with a Lorentz metric $\Tilde{g}$. A key aspect of such modeling is that it presupposes the choice of unit length. However, if different observers agree to measure lengths at different points with different units, then not only one Lorentz metric $\Tilde{g}$ is being used but a whole set of them. In turn, this set is a class of equivalence $\mathcal{G}$ of metrics $g$ such that $g = e^\lambda \Tilde{g}$, with $\lambda$ being an arbitrary smooth function on $M$. Moreover, a Weyl structure on $M$ is a map $F : \mathcal{G} \rightarrow \Lambda^1(M)$, where $\Lambda^1(M)$ is the space of 1-forms on $M$, defined by $F(e^\lambda g) = F(g) - d\lambda$, whilst a manifold with a Weyl structure is called a Weyl manifold \cite{Folland2008tourist}. One of the key aspects of this geometrical approach is that Weyl structures allow us to compare lengths of vectors at different points of $M$ through the transport of scalar products, associated to the metrics in $\mathcal{G}$, along curves over the manifold \cite{Folland1970weyl}.
	
	Furthermore, such theory enables us to understand the gauge invariance principle in a full geometrical perspective, see \cite{Folland2008tourist} for more details. In particular, it reveals how the commonly seen ``covariant derivative" $D_\mu = \partial_\mu + iA_\mu$ appears in the context of field theories. Nevertheless, such procedure to define Weyl structures is not the only one, as it can be seen in \cite{Hall1992}.
	
	The central role played by local invariance is not only restricted to (quantum) field models. Therefore, it is natural to step back and seek its implications on systems with finite degrees of freedom. In effect, there is a long list of relevant models where local or gauge invariance is as important as described previously. For example, the dynamics of any classical-mechanics system can be formulated with a local structure \cite{Debbasch1993, Deriglazov_Rizzuti2011}. We have only to take into account the corresponding reparametrization-invariant formulation. In this case, reparametrization invariance (RI) can be seen as a gauge symmetry \cite{Fulop_1999}. A mathematical oriented perspective on all that can be seen in \cite{Massa2000}. 
	
	RI has even deeper implications though. It may lead to a common arrow of time and non-negativity of mass \cite{gueorguiev_reparametrization_2021}. Broader consequences, such as an underlying origin to classical long range interactions (electrodynamics and gravitation) could also be supported by reparametrization-invariant systems \cite{gueorguiev_geometric_2021}. Gauge theory for finite-dimensional dynamical systems has also been explored, running from quantum mechanics to information theory, passing through chemistry and rigid body dynamics \cite{Gurfil2007}. 
	
	In a more philosophical approach, in the realm of Newtonian physics, one could eliminate the notion of an absolute space by gauging translations and rotations \cite{Ferraro2016}. While this is restricted to the Galileo group, we extend it in this paper by considering the gauge principle directly applied to arbitrary Lagrangian systems with finite degrees of freedom. We will name it \textit{Classical Gauge Principle}. The prescription is pretty much the same: a Lagrangian model inert under the representation of a Lie group parameterized by global parameters is reformulated by the swap of such parameters to local ones, that is, smooth functions of the evolution variable. Just as before, it implies the introduction of a covariant derivative (and the corresponding connection) into the original model.
	
	The paper is divided as follows. In Section \ref{sec.free_dirac} we present a seminal and illustrative example which can be rewritten in order to present local invariance, according to the gauge principle. This example is a key application of such principle and, in turn, the consequences of its usage are then enumerated. In Section \ref{introducing.local.invariance} we develop the method of introducing local invariance in classical-mechanics models and present two modern examples in which this technique may be applied. In Section \ref{local.structure} we formalize the concept of equivalent classes of Lagrangians with the same local structure. Furthermore, leveraging on a more pragmatic view, two examples are given to elucidate such approach. Finally, Section \ref{sec.concl} is left for the conclusions.
	
	\section{Free Dirac Lagrangian}\label{sec.free_dirac}
	
	Let us consider an underlying example of a model endowed with local invariance. We start with the Lagrangian of a free massive spinor field $\psi = \psi(x^\mu)$ and write
	
	\begin{equation}
		\mathcal{L} = \Bar{\psi}(i\gamma^\mu\partial_\mu - m)\psi.\label{Dirac}
	\end{equation}
	Here, $\Bar{\psi} = \psi^\dagger\gamma^0$, whereas $\gamma^\mu$ are the Dirac-gamma matrices and $\gamma^\mu\gamma^\nu + \gamma^\nu\gamma^\mu = 2\eta^{\mu\nu}$, with $\eta^{\mu\nu} = \text{diag}(1,-1,-1,-1)$. In turn, $\mathcal{L}$ is invariant under the action of the symmetry group $U(1)$, since the transformations $\psi \mapsto \psi' = e^{i\alpha}\psi$ and $\Bar{\psi} \mapsto \Bar{\psi}' = e^{-i\alpha}\Bar{\psi}$, where $\alpha$ is a real number, do not alter the Lagrangian,
	\begin{equation}
		\mathcal{L} \mapsto \mathcal{L}'=  e^{-i\alpha}\Bar{\psi}(i\gamma^\mu\partial_\mu - m)e^{i\alpha}\psi = \mathcal{L}.
	\end{equation}
	One may ask now if this Lagrangian would still be invariant under transformations where $\alpha$ is an arbitrary smooth function. Naturally, it would no longer be invariant under the action of the same transformations, due to the derivative $\partial_\mu\psi$ term. However, local invariance can be achieved in this scenario. In order to enforce such property, we introduce the vector field $A_\mu$ in $\mathcal{L}$ and define
	\begin{equation}
		D_\mu := \partial_\mu - iA_\mu.
	\end{equation}
	We also fix the transformation of $A_\mu$ with
	\begin{equation}
		A_\mu \mapsto A'_\mu = A_\mu + \partial_\mu\alpha.\label{ex1_transf_Amu}
	\end{equation}
	This way, we write
	\begin{equation}
		\mathcal{L} = \Bar{\psi}(i\gamma^\mu D_\mu - m)\psi,
	\end{equation}
	whereas $D_\mu\psi$ has a simple transformation law,
	\begin{equation}
		D_\mu\psi \mapsto D'_\mu\psi' = e^{i\alpha(x)}D_\mu\psi.
	\end{equation}
	Therefore, local invariance for $\mathcal{L}$ has been established, as previously claimed. Such procedure is the standard way in order to obtain the interaction of the initial free field with the vector $A_\mu$, commonly called the gauge field.
	
	Let us make 
	two elucidating observations.
	\begin{enumerate}
		\item Models with local invariance are constructed according to this recipe, known as the gauge principle. Namely, a field model which is invariant under a Lie group of global transformations may be reformulated to be invariant under the same group, although with local parameters. In order to do so, an interaction with a vector field is introduced, whilst the components of this field, named gauge field, are but the components of a 1-form which defines the connection map on a Weyl manifold \cite{Folland2008tourist}.
		
		\item Lagrangian models with local invariance present first class constraints through the corresponding hamiltonization procedure \cite{aad.rizzuti.symmetries.PhysRevD.83.125011, guilherme.matheus.rizzuti.2020}. Theories of this type possess spurious degrees of freedom, with ambiguous evolution. The physical sector may be consistently characterized by gauge invariant variables.
	\end{enumerate}
	
	In our illustrative example, $A_\mu$ is the vector potential. Since it changes under the local transformation, see \eqref{ex1_transf_Amu}, $A_\mu$ has no physical interpretation. The possible observables of the model are combinations that are not affected by the local transformations; for instance, the integral $I = \oint\Vec{A}.d\Vec{l}$ of the spatial part of $A_\mu$ over any closed contour is invariant under \eqref{ex1_transf_Amu}. It manifests itself through the Aharonov-Bohm effect \cite{aharonov.bohm.PhysRev.115.485}. The electromagnetic tensor, represented by $F_{\mu\nu} = \partial_\mu A_\nu - \partial_\nu A_\mu$, is another example of a local invariant which is an observable.
	
	At this point, a natural question that rises is if the gauge principle for field models can be transposed to systems with finite degrees of freedom. The answer is yes and the Classical Gauge Principle will be discussed in details in the next section, followed by applications in both classical and relativistic mechanics.
	
	\section{Introducing local invariance in classical mechanics models}\label{introducing.local.invariance}
	
	Local invariance is not a privilege of field theoretical models. Actually, in classical mechanics of constrained systems it is also present, which leads us to the so-called Dirac conjecture \cite{dirac_lectures_2001}. It states that primary first class constraints (FCC) are generators of transformations of configuration and momenta variables, keeping the physical state of the system untouched. That is not the whole story though. In references \cite{gitman_quantization_1990, henneaux_quantization_1992}, we may find a sequence of theorems connecting local symmetries and first class constraints. To sum it up:
	
	\textbf{i.} In a singular Lagrangian model, the solution of equations of motion contains as many arbitrary functions of time as the number of primary FCC present throughout the corresponding hamiltonization. 
	
	\textbf{ii.} If a singular Lagrangian is invariant under a local transformation parame-trized by arbitrary functions and their derivatives up till order $\kappa \in \mathbb{N}$, then, in general, there are constraints of $\kappa +1$ stage. 
	
	\textbf{iii.} There are $n \in \mathbb{N}$ independent identities fulfilled by the Lagrangian equations of motion if, and only if, there exist symmetries of the Lagrangian action with $n$ parameters, which are arbitrary functions of time. 
	
	In order to illustrate the aforementioned results, let us consider a toy model. It is defined on configuration space with coordinates $q^A \equiv (x,y,z)$ and the Lagrangian action is given by 
	\begin{equation}\label{action}
		S = \int d \tau \left [ \frac{1}{2}(\dot x - y)^2 +\frac{1}{2} (z+\dot y)^2 \right ].
	\end{equation}
	As usual, $\dot q^A \equiv \frac{dq^A}{d\tau}$ and $\tau$ is the time evolution parameter. This model is but a slight variation of an example presented in \cite{gitman_quantization_1990}. The equations of motion read,
	\begin{equation}\label{94}
		\frac{\delta S}{\delta x}=0 \Rightarrow \ddot x-\dot y=0,
	\end{equation}
	\begin{equation}\label{95}
		\frac{\delta S}{\delta y}=0 \Rightarrow \dot z+ \ddot y=-(\dot
		x-y),
	\end{equation}
	\begin{equation}\label{96}
		\frac{\delta S}{\delta z}=0 \Rightarrow z+ \dot y=0.
	\end{equation}
	
	By direct inspection, we find the identity
	\begin{equation}\label{97}
		\left (\frac{d^2}{d\tau^2}\frac{\delta}{\delta
			z}+\frac{d}{d\tau}\frac{\delta}{\delta y}-\frac{\delta}{\delta
			x} \right )S=0.
	\end{equation}
	It expresses the fact that the action \eqref{action} is invariant under the local transformations
	\begin{equation}\label{98}
		x\rightarrow x'=x+\al,
	\end{equation}
	\begin{equation}\label{99}
		y \rightarrow y'=y+\dot \al,
	\end{equation}
	\begin{equation}\label{100}
		z \rightarrow z'=z-\ddot \al.
	\end{equation}
	As stated in comment \textbf{iii}, this symmetry is parametrized by only one parameter, the arbitrary function of $\tau$ named $\alpha$. 
	
	If we now fix initial conditions $x(0)= x_0$, $\dot x(0) = y_0$, $y(0) = y_0$, $\dot y(0) = -z_0$ and $z(0) = z_0$, we find the solutions
	\begin{equation}\label{106}
		x=x_0+y_0 \tau -\frac{z_0}{2}\tau^2-\int_0^{\tau}\left (\int_0^{\tau
			'}\varphi(\tau '')d\tau '' \right )d \tau ' ,
	\end{equation}
	\begin{equation}\label{107}
		y=y_0-z_0\tau - \int_0^{\tau}\varphi(\tau')d \tau',
	\end{equation}
	\begin{equation}\label{108}
		z=z_0+ \varphi(\tau),
	\end{equation}
	where $\varphi$ is an arbitrary function (excpet that $\varphi(0)=0$). Clearly, it is a constrained system. The corresponding Hamiltonian formulation can be constructed with the Dirac algorithm \cite{deriglazov_classical_2017}. The canonical momenta $p_A =\frac{\partial L}{ \partial \dot q^A}$ read
	\begin{equation}\label{109}
		p_x=\frac{\partial L}{\partial \dot x}=\dot x-y \Rightarrow \dot x
		=p_x+y,
	\end{equation}
	\begin{equation}\label{110}
		p_y=\frac{\partial L}{\partial \dot y}=z+ \dot y \Rightarrow \dot
		y=p_y-z,
	\end{equation}
	\begin{equation}\label{111}
		p_z=\frac{\partial L}{\partial \dot z}=0 \,\,\,  \mbox{(Primary constraint)}.
	\end{equation}
	The Hamiltonian $H_0$ and total Hamiltonian $H$ are given  by
	\begin{equation}\label{112}
		H=H_0+\lambda p_z,
	\end{equation}
	where,
	\begin{equation}\label{113}
		H_0=\frac{1}{2}p_x^2+\frac{1}{2}p_y^2+yp_x-zp_y
	\end{equation}
	and $\lambda$ is the Lagrange multiplier for the primary constraint $p_z = 0$. 
	Defining the Poisson brackets
	\begin{equation}
		\{ A,B \} = \frac{\partial A}{\partial q^A}\frac{\partial B}{\partial p_A} - \frac{\partial A}{\partial p_A}\frac{\partial B}{\partial q^A}, 
	\end{equation}
	we may proceed to the further stage constraints. The consistency condition $\dot p_z = 0$ implies the secondary constraint
	\begin{equation}\label{114}
		0=\{p_z,H\}=p_y \Rightarrow p_y=0.
	\end{equation}
	Also, $\dot p_y = 0$ provides
	\begin{equation}\label{115}
		0=\{p_y,H\}=-p_x \Rightarrow p_x=0,
	\end{equation}
	Since $\dot p_x = 0$ brings no new information, the procedure stops here. Since all the constraints commute with each other, they are all FCC. It is not a surprise that the procedure stops at the third step: the $\ddot \alpha$-symmetry in the Lagrangian formulation would provide a tertiary constraint; see comment \textbf{ii}. Furthermore, we point out that the Lagrange multiplier $\lambda$ hasn't been found in the course of the procedure. In this case, it enters into the equation of motion  
	\begin{equation}
		\dot z = \lambda
	\end{equation}
	as an arbitrary function, in accordance with \eqref{108}. 
	
	To conclude these initial remarks, let us see how the primary FCC generates the local symmetry already presented in \eqref{98}-\eqref{100}. We start with the variable $z$. For an infinitesimal time lapse $\delta \tau$ \cite{dirac_lectures_2001},
	\begin{align}
		z(\delta \tau) &= z(0) + \dot z \delta \tau = z(0) + \delta \tau \{z, H \} \nonumber \\
		&= z(0) + \delta \tau \{z, H_0 \} + \delta \tau \lambda \{z, p_z \}.
	\end{align}
	We could, however, take another multiplier due to its arbitrariness,
	\begin{equation}
		z'(\delta \tau ) = z(0) + \delta \tau \{z, H_0 \} + \delta \tau \lambda' \{z, p_z \}.
	\end{equation}
	That is, 
	\begin{equation}\label{deltaz}
		\Delta z \equiv z'-z = \delta \tau (\lambda' - \lambda)\{z, p_z \}.
	\end{equation}
	For convenience, we could rename the arbitrary function $\delta \tau (\lambda' - \lambda)$ by $-\ddot \alpha$. This way, the transformation law \eqref{deltaz} is reduced to \eqref{100}, showing that the FCC $p_z=0$ is indeed the corresponding generator. The transformations for $x$ and $y$ follow from the very structure of equations of motion. 
	
	Based on the Dirac conjecture and the deep liaison between FCC and local transformations, many works have attacked the problem of obtaining  constructively all the gauge symmetries of a singular Lagrangian model, see for example \cite{borochov.tyutin, gitman_symmetries_2006, deriglazov_reconstruction_2007} and references therein.  
	
	So far, the local symmetries were presented within the Lagrangian structure. Is there a possible constructive manner to impose them, mimicking what is usually done with the gauge principle in field theories? The answer is, in fact, yes and we will demonstrate it throughout the remaining part of this Section.   
	
	To begin with, let us consider a configuration space parameterized by ${x^a}$, $a=1,\dots,n$ and a mechanical model defined on it. The corresponding dynamics is governed by a Lagrangian $L(x^a,\dot{x}^a)$, where $\dot{x}^a:=\frac{dx^a}{d\tau}$ and $\tau$ is an evolution parameter. As usual, the equations of motion are obtained from a least action principle applied to the action functional
	\begin{equation}
		S = \int d\tau L(x^a,\dot{x}^a).
	\end{equation}
	
	We also impose that $L$ is invariant under the faithful  representation of a matrix Lie group $\mathcal{G}$, whose elements are denoted by $G$, with respective entries $G^{a}_{\;\;b}$ in a field $\mathbb{F}$. The linear representation is defined by
	\begin{equation}
		x^a \rightarrow x'^a := G^{a}_{\;\;b}x^b.
	\end{equation}
	Although it appears to be a restriction to consider only $n\times n-$matrix groups directly realized onto the underlying ($n$-dimensional) configuration space, every case of interest to this work possess this very structure. We could go even further and realize that there are many standard examples where this particular representation plays a central role, for instance $SO(3)$ as the rotation group in space or even $SO(1,3)$ describing the Lorentz transformations in spacetime and so on \cite{adb}.
	
	The corresponding Lie algebra is denoted by 
	\begin{equation}
		\mathfrak{g} = \{\xi^i\Gamma_i\;:\; \xi^i \in \mathbb{F},\; [\Gamma_i,\Gamma_j] = c_{ij}^{\;\;\;k}\Gamma_k\},
	\end{equation}
	that is, $\mathfrak{g}$ is generated by $\Gamma_i, i=1,\dots,n$ and $c_{ij}^{\;\;\;k}$ are the structure constants. Any $G \in \mathcal{G}$ sufficiently close to the identity of the group may be written as
	\begin{equation}
		G = e^{\xi^i\Gamma_i},
	\end{equation}
	with $G^{a}_{\;\;b} = ( e^{\xi^i\Gamma_i})^{a}_{\;\;b}$. It will be sufficient to this work to write the group elements only in first order in the exponential map,
	\begin{equation}
		G^{a}_{\;\;b} \approx \delta^{a}_{\;\;b} + \xi^i(\Gamma_i)^{a}_{\;\;b}.
	\end{equation}
	According to the initial assumption,
	\begin{align}
		L(x^a,\dot{x}^a) \xrightarrow{G} L'(x'^a,\dot{x}'^a) &= L(G^{a}_{\;\;b}x^b,G^{a}_{\;\;b}\dot{x}^b)\nonumber\\ 
		&= L(x^a,\dot{x}^a).\label{transfo_L}
	\end{align}
	
	Now, we may relax this first assumption to include not only constant parameters $\xi^i$, but also smooth functions
	\begin{align}
		\xi^i:I \subset \mathbb{R} &\rightarrow \mathbb{F},\;\forall\;i\in\{1,\dots,n\}\\
		\tau &\mapsto \xi^i(\tau),\nonumber
	\end{align}
	where $I$ is an open interval of the real line. We denote both the Lie algebra and group so obtained by $\mathfrak{g}_\tau$ and $G_\tau$. It means they are local in the sense that their elements are parameterized by arbitrary smooth functions of $\tau$. Clearly, due to the velocity $\dot{x}^a$ in $L$, the latter is no longer invariant under the $G-$action,
	\begin{align}
		\frac{dx^a}{d\tau} \rightarrow \frac{dx'^a}{d\tau} &= \frac{d}{d\tau}(\delta^a_{\;\;b} + \xi^i(\Gamma_i)^a_{\;\;b})x^b\\
		&= G^a_{\;\;b}\dot{x}^b + \dot{\xi}^i(\Gamma_i)^a_{\;\;b}x^b.\label{transfo_xdot}
	\end{align}
	
	We would like to maintain its invariance even under the action of the local group $G_\tau$. Hence, we proceed exactly in the same manner one introduces covariant derivatives through the gauge principle in field theories. In our cause, though, we substitute the time derivative for
	\begin{equation}
		\delta^a_{\;\;b}\frac{d}{d\tau} \rightarrow D^a_{\;\;b} := \delta^a_{\;\;b}\left(\frac{d}{d\tau} - g\right),
	\end{equation}
	where $g$ is a new variable, analogous to the gauge field. The Lagrangian remains invariant under $G_\tau$, provided $g$ changes according to
	\begin{equation}
		\delta^a_{\;\;b}g \rightarrow \delta^a_{\;\;b}g' = \delta^a_{\;\;b}g + \dot{\xi}^i(\Gamma_i)^a_{\;\;b}.\label{delta_g}
	\end{equation}
	In fact, the pathological term comes from the velocity, see \eqref{transfo_xdot}. However, it is ruled out by the transformation provided above in \eqref{delta_g},
	\begin{equation}
		D^a_{\;\;b}x^b \rightarrow {D'^a}_{b}x'^b =  G^a_{\;\;b} D^b_{\;\;c}x^c.
	\end{equation}
	In turn, returning back to $L$,
	\begin{align}
		L(x^a,D^a_{\;\;b}x^b) \rightarrow L(x'^a,D'^a_{\;\;b}x'^b) &= L(G^a_{\;\;b}x^b,G^a_{\;\;b}D^b_{\;\;c}x^c)\nonumber\\
		&= L(x^a,D^a_{\;\;b}x^b),
	\end{align}
	where the equality comes from \eqref{transfo_L}.
	
	As claimed previously, we have imposed local invariance to the action we are working with. In the next subsections, we apply the procedure developed here to concrete models, revealing the power of the method while explaining a possible geometrical and algebraic origin to leading proposals, such as the free relativistic particle, as well as more sophisticated examples, such as doubly special relativity (DSR) models and spinning particles.
	
	\subsection{Free particle and DSR proposal}
	
	In order to discuss our next particle model, let us introduce a motivation to study DSR models. Advances in the development of a quantum theory of gravity suggest the existence of a fundamental scale \cite{amelino-camelia_relativity_2002} that would imply in a discrete spectrum for measurable quantities such as areas and volumes \cite{rovelli_discreteness_1995}. However, as a consistency condition it is expected that such theory would agree with the theory of special relativity in the regime in which gravitational effects may be neglected. In turn, doubly special relativity flourished as a theory attempting to reformulate special relativity through the insertion of a new invariant scale, besides the speed of light, while keeping an intermediate regime with fixed spacetime as background. A complete discussion on facts, myths and open questions about DSR may be found in \cite{amelino-camelia_doubly-special_2010}.

	The model in \cite{magueijo_lorentz_2002}, by J. Magueijo and L. Smolin, is one of the most accepted DSR proposals up to date. It is built upon the idea that the dispersion relation for the conserved four-momentum of a particle should be written as
	\begin{equation}
		p_\mu p^\mu = m^2c^2(1 + \xi p^0)^2,\label{dispersion_MS}
	\end{equation}
	accompanied by the non-linear representation of the Lorentz group
	\begin{equation}
		p'^{\mu} = \frac{\Lambda^\mu_{\;\nu} p^\nu}{1 + \xi(p^0 - \Lambda^0_{\;\nu}, p^\nu)},\label{momentum_lorentz}
	\end{equation}
	responsible for maintaining \eqref{dispersion_MS} unaltered. Here $\xi$ represents the invariant scale of the model, related to the Planck length. Moreover, the quantity $p^\mu = \left(-\frac{1}{\xi},0,0,0\right)$ is kept intact under \eqref{momentum_lorentz}. Nonetheless, such proposal is not without issues. For instance, it lacks a fully covariant space-time description, due to its construction on the energy-momentum space. In fact, recent results \cite{DSR.PLB.Guilherme.2021} have shown that it may be seen as a particular gauge of a free particle model. Furthermore, using a four-dimensional spacetime as a cornerstone to the construction of Lagrangian DSR models is a delicate topic \cite{rizzuti_comment_2010}. With that in mind, we shall present a particle model built according to the Classical Gauge Principle that results in the aforementioned free particle model. We emphasize that the introduction of local invariance in Lagrangian actions as proposed here leads to the presence of FCC on the corresponding Hamiltonian formulation. In that case, we are allowed to fix the corresponding gauges, producing, in our example, the so desired deformed dispersion relation \eqref{dispersion_MS}.

	Let us consider 
	a particle model described by the Lagrangian
	\begin{equation}
		L = \frac{1}{2}m\eta(v,v),\label{lagrangian_freep}
	\end{equation}
	defined on $(\mathbb{R}^5,\eta)$, a 5-dimensional spacetime endowed with a pseudo-metric $\eta$ and parameterized by $\{x^A\}$. $m$ is a constant that can be interpreted as the mass of the particle.  We shall denote $v^A = \dot{x}^A$ and write $\eta_{AB}=\text{diag}(+1,-1,-1,-1,-1)$. We also allow ourselves to multiply position and velocity vectors by complex numbers and impose
	\begin{equation}
		\eta(\alpha v_1,\beta v_2) = (\alpha^*\beta)\eta(v_1,v_2),\;\forall\;\alpha,\beta \in \mathbb{C}.
	\end{equation}
	With these considerations, we take the representation of the group $U(1)=\{e^{i\gamma};\;\gamma \in \mathbb{R}\}$ on $(\mathbb{R}^5,\eta)$ given by
	\begin{equation}
		x^A \rightarrow x'^A = e^{i\gamma}x^A \Rightarrow v'^A = e^{i\gamma}v^A,
	\end{equation}
	which gives us
	\begin{align}
		L \rightarrow L' = \frac{1}{2}m\eta(v',v') &= \frac{1}{2}m\eta(e^{i\gamma}v,e^{i\gamma}v)\\
		&= \frac{1}{2}m\eta(v,v) = L.
	\end{align}
	
	According to the classical gauge principle, we raise $U(1)$ to the status of a local group, meaning that its elements are now parameterized by smooth functions $\gamma : I\; (\text{open}) \subset \mathbb{R} \rightarrow \mathbb{R}$. As explained, $L$ is no longer invariant under $x'^A = e^{i\gamma(\tau)}x^A$. In accordance with what we have exposed previously, we substitute
	\begin{equation}
		\frac{d}{d\tau} \rightarrow D:= \frac{d}{d\tau} -i\gamma.
	\end{equation}
	This way, in local coordinates the Lagrangian reads
	\begin{align}
		L &= \frac{1}{2}m\eta_{AB}Dx^ADx^B\\
		&= \frac{1}{2}m\eta_{AB}(\dot{x}^A - i\gamma x^A)(\dot{x}^B - i\gamma x^B),\label{base_Lagrangian_free_particle}
	\end{align}
	which is now invariant under $U(1)_\tau = \{e^{i\gamma(\tau)};\gamma : I \rightarrow \mathbb{R}\}.$ If we redefine $i\gamma = g$, we obtain exactly the same action proposed in \cite{DSR.PLB.RIZZUTI.2011}. Due to the local invariance, the Lagrangian presents first class constraints through the Dirac's hamiltonization algorithm \cite{dirac_lectures_2001}. The physical sector of the model describes a free massive particle and the superior bounded speed is obtained in a particular gauge choice.
	
	We conclude this section with a brief comment concerning the transformation law given in \eqref{delta_g}. It implies that the gauge variable $g$ changes as a total derivative,
	\begin{equation}
		\Delta(\delta^a_{\;\;b}g) := \delta^a{}_b(g'-g) = \frac{d}{d\tau}\left(\xi^i(\tau)(\Gamma_i)^a_{\;\;b}\right).
	\end{equation}
	In this case, we could change $L$ in \eqref{lagrangian_freep} for
	\begin{equation}
		L \rightarrow L' = L - \xi g = \frac{1}{2}m\eta Dx^ADx^B - \xi g,\label{lagrangian_DSR}
	\end{equation}
	where $\xi$ is a constant that couples the gauge variable $g$ to the dynamical sector of the model. In turn, it can be shown that these Lagrangians are equivalent and possess the same local structure (this notion shall be clarified in Section \ref{local.structure}). Moreover, \eqref{lagrangian_DSR} is the Lagrangian described in the work \cite{DSR.PLB.Guilherme.2021}, in which it was shown that the MS DSR model is a particular gauge of a free relativistic particle Lagrangian model. In this particular case, $\xi$ is related to the new invariant scale, present within the model from the beginning. In fact, the transition to the Hamiltonian formulation starts with the conjugate momenta
	\begin{align}
		p_A &= \frac{\partial L}{\partial\dot{x}^A} = m\eta_{AB}(\dot{x}^B - gx^B),\label{momenta_pA}\\
		p_g &= \frac{\partial L}{\partial\dot{g}} = 0.\label{momenta_g}
	\end{align}
	The Hamiltonian assumes the form
	\begin{align}
		H(Y^a, p_a) &= \left(p_a\dot{Y}^a - L\right)\vert_{\eqref{momenta_pA},\eqref{momenta_g}} + v_gp_g  \nonumber \\
		&= \frac{1}{2m}\eta^{AB}p_Ap_B + g\eta^{AB}p_Ax_B + \xi g + v_gp_g\label{complete_hamiltonian}
	\end{align}
	where $v_g$ is the Lagrange multiplier for the constraint $p_g=0$ and we are compressing the configuration and momenta variables into the notation $(Y^a, p_a)$. The vanishing condition of the constraints throughout all the time implies the following chain, 
	\begin{align}
		\dot p_g = 0 \Rightarrow p_Ax^A + \xi = 0. \\ 
		(p_Ax^A + \xi)\dot{}=0 \Rightarrow \eta^{AB}p_Ap_B = 0.
	\end{align}
	The procedure stops at this stage, since the time evolution of $\eta^{AB}p_Ap_B = 0$ brings no new information. All the constraints are of first class. By fixing the gauge condition 
	\begin{equation}
		p^5 = mc(1 + \xi p^0)\label{MS_DSR_gauge}
	\end{equation}
	to the constraint $p_Ax^A + \xi = 0$, they form a second class pair and as such can be eliminated from the description. Also, the direct substitution of this gauge back to the constraint $\eta^{AB}p_Ap_B = 0$ provides the desired MS DSR dispersion relation, as claimed. The non-linear transformation law for the four-momenta may also be found by imposing local invariance to the gauge \eqref{MS_DSR_gauge}. In order to not extend too much our considerations, all the details concerning the previous analysis can be found in the reference \cite{DSR.PLB.Guilherme.2021}. 
	
	This example reveals the power of the Classical Gauge Principle, allowing us to arrive at concrete models from a somewhat generic Lagrangian.
	
	\subsection{Classical spinning particle}
	
	Although the proper description of spin could only be achieved in the realm of quantum electrodynamics \cite{alexei_rizzuti_genaro_2012}, many attempts have been done in order to consistently fathom it (semi) classically, see, for example, \cite{deriglazov_non-grassmann_2012, arxiv.1909.07170} and references therein. The first proposals on this path can be traced back to the works of Frenkel \cite{frenkel_elektrodynamik_1926} and Thomas \cite{thomas_motion_1926}. Also, Bargmann, Michel and Telegdi have shown that their model almost reproduces the proper dynamics of spin in uniform fields \cite{bargmann_precession_1959}. Naturally, one expects these models to produce Dirac's equation (or the Pauli one, depending on the corresponding energy) upon quantization, which unfortunately was not the case. Furthermore, the seminal papers of Berezin and Marinov partially filled this blank \cite{berezin_marinov_1975, berezin_particle_1977}. While their model provided the Dirac equation when quantized, yet it was formulated with Grassmann variables, which could lead to certain difficulties in classical regimes \cite{berezin_particle_1977}. 
	
	Curiously enough, these models present interesting quantum features, even before quantization. For example, in \cite{rizzuti_electron_2014}, the angular momentum associated to the Zitterbewegung\footnote{The Zitterbewegung is a trembling motion the free electron would experience. It was first predicted by Schrödinger when analyzing the Dirac equation \cite{sch.zitter.1930}.} can only possess two orientations and is bounded from above by $\hbar/2$.  
	
	We turn our attention to a particular model of spinning particle \cite{Deriglazov_2010.sigma}. It will be interesting to our proposal here as we can impose local invariance. Contrary to our previous example, the covariant derivative encodes the interaction of the model with an external field. Moreover, the local parameters gain clear physical interpretation, as we shall see. 
	
	On the configuration space parameterized by $\{v^i, g, \varphi \}, \, i \in \{1,2,3\}$, we define the following Lagrangian, 
	\begin{equation}
		L_{spin} = \frac{1}{2g}(\dot v^i)^2 +g \frac{b^2}{2a^2} + \frac{1}{\varphi}[(v^i)^2-a^2].\label{lagrangian_spin}
	\end{equation}
	We are using the short hand notation $(v^i)^2 = \delta_{ij}v^i v^j = v^2$, as well as $\dot v^i = \frac{dv^i}{d \tau}$. On one hand $a$ is an arbitrary constant, while $b^2 = 3 \hbar^2 /4$, whose value will be clarified in a while. 
	
	Let us discuss the dynamics provided by \eqref{lagrangian_spin}. First of all, the Lagrangian is invariant under the local transformations
	\begin{equation}\label{symmetries.1}
		\begin{split}
			\delta v^i = \alpha \dot v^i,  \\
			\delta g = (\alpha g)\dot{}, \\
			\delta \varphi  = \alpha \dot \varphi - \dot \alpha \varphi,
		\end{split}
	\end{equation}
	with $\alpha = \alpha(\tau)$ is an arbitrary function. According to our previous discussions, the presence of gauge symmetries indicates the existence of constraints on the model. We will reveal them during the hamiltonization carried out with the Dirac method. In the first place, one defines the conjugate momenta, which are used as algebraic equations to find the velocities in terms of configuration and momenta variables,
	\begin{equation}\label{velocities}
		\begin{split}
			\pi_i &= \frac{\partial L_{spin}}{\partial \dot v^i}= \frac{\dot v_i}{g} \Rightarrow \dot v_i = g \pi_i, \\
			\pi_g &= \frac{\partial L_{spin}}{\partial \dot g} = 0, \\
			\pi_\varphi &= \frac{\partial L_{spin}}{\partial \dot g} =0.
		\end{split}
	\end{equation}
	If we collect all the configuration (momenta) variables under $q^A$ ($\pi_A$), then the Hamiltonian $H_0$ and total Hamiltonian $H$ are defined by
	\begin{equation}
		H = H_0 + \lambda_g \pi_g + \lambda_\varphi \pi_\varphi
	\end{equation}
	where
	\begin{align}
		H_0 &= (\pi_A \dot q^A - L_{spin})\vert _{\mbox{\footnotesize \eqref{velocities}}} \nonumber \\
		&= \frac{g}{2}\left (\pi^2 - \frac{b^2}{a^2} \right )-\frac{1}{\varphi}\left (v^2-a^2 \right ).
	\end{align}
	
	Following up with the Dirac prescription, one defines the Poisson brackets for two functions on the phase space
	\begin{equation}
		\{F, G \} = \frac{\partial F}{\partial q^A}\frac{\partial G}{\partial \pi_A} - \frac{\partial F}{\partial \pi_A}\frac{\partial G}{\partial q^A};
	\end{equation}
	the time evolution of any function stands as $\dot A = \{A, H \}$. With this structure in hands, we impose the consistency conditions on both primary constraints, generating superior stage ones,
	\begin{equation}
		\begin{array}{cc}
			\dot \pi_g = 0 \Rightarrow \pi^2 = \frac{b^2}{a^2}  \\
			\dot \pi_\varphi = 0 \Rightarrow v^2 = a^2 \\
			\pi \dot \pi = 0 \Rightarrow v \pi = 0 \\
			(v \pi)\dot{} = 0 \Rightarrow \frac{2}{\varphi}a^2 + g \frac{b^2}{a^2}=0.  
		\end{array}
	\end{equation}
	The auxiliary sector $(g, \varphi)$ is restricted to the second class constraints (SCC) $\pi_\varphi=0$ and $\frac{2}{\varphi}a^2 + g \frac{b^2}{a^2}=0$, whilst the FCC $\pi_g=0$ is related to the gauge symmetry \eqref{symmetries.1}. The gauge $g=1$ forms a second class pair with with $\pi_g=0$, that allows us to omit all the $(g, \varphi)$ part. We are left with a constraint surface defined by 
	\begin{equation}
		\begin{split}
			v^2=a^2, \quad  v \pi=0 \quad \mbox{(SCC)}, \\
			\pi^2 = \frac{3\hbar^2}{4a^2} \quad \mbox{(FCC)}.
		\end{split}
	\end{equation}
	
	While the variables $v^i$ and $\pi_i$ are explicitly changed by the symmetries \eqref{symmetries.1}, we may find a combination of them that remains intact under these transformations. We take 
	\begin{equation}\label{j1}
		J_i = \varepsilon_{ijk}v^j \pi^k
	\end{equation}
	for distinct reasons. Firstly, it is invariant under \eqref{symmetries.1}, $\delta J_i = 0$. This result promotes $J_i$ as a possible observable within the theory. Secondly, due to the constraint structure, we have
	\begin{equation}\label{j2}
		J^2 = v^2\pi^2 - (v \pi)^2 = \frac{3\hbar^2}{4}.
	\end{equation}
	At last, we also observe that
	\begin{equation}\label{j3}
		\{J_i, J_j \} = \varepsilon_{ijk} J^k.
	\end{equation}
	Clearly, \eqref{j3} satisfies the standard $SO(3)$ algebra, while \eqref{j2} resembles a finite dimensional irreducible representation of angular momentum operators when labeled by $s=1/2$. Thus, we quantize the variables $J_i$ according to 
	\begin{equation}\label{quantization}
		J_i \longrightarrow \hat{J}_i = \frac{\hbar}{2} \sigma_i,
	\end{equation}
	where $\sigma_i$ are the Pauli matrices. 
	
	Although lengthy, our previous exposition makes it clear why the presented model is particularly important to describe internal degrees of freedom of classical spinning particles.  
	
	Now we shall focus on \eqref{lagrangian_spin}, as another application to the Classical Gauge Principle developed here. It is clear that this Lagrangian has manifest $SO(3)$ global invariance,
	\begin{equation}
		v^i \to v'^i = {R^i}_jv^j;\;\;\;\forall\;R\in SO(3).
	\end{equation}
	Identifying $SO(3)$ as a matrix Lie group we may write
	\begin{equation}
		{R^i}_j = {(e^{\xi^k\Gamma_k})^i}_j,
	\end{equation}
	where $\xi^k\Gamma_k$ is an element of the Lie algebra $\mathfrak{so}(3)$ and $\xi^k$ are interpreted as rotation angles. According to our prescription, we promote the constant parameters $\xi^k$ to arbitrary smooth functions $\xi^k(\tau)$. In order to maintain the invariance of $L_{spin}$, we introduce the covariant derivative
	\begin{equation}
		{\delta^i}_j\frac{d}{d\tau} \to {\delta^i}_j\frac{d}{d\tau} + \dot{\xi}^k{(\Gamma_k)^i}_j.
	\end{equation}
	In this case $\dot{\xi}^k$ may be interpreted as an angular velocity. We redefine it according to
	\begin{equation}
		\dot{\xi}^k = -\frac{e}{m}B^k,
	\end{equation}
	which provides the angular velocity of a charged massive particle in terms of an external magnetic field.   
	
	Together with this redefinition, there is an explicit form for the generators $\Gamma$,
	\begin{equation}
		{(\Gamma_k)^i}_j = \ve_{ikj},
	\end{equation}
	where $\ve_{ijk}$ is the Levi-Civita symbol. Thus, \eqref{lagrangian_spin} may be rewritten as
	\begin{equation}
		L'_{spin} = \frac{1}{2g}(\dot v^j - \frac{e}{m}\ve_{jik}v^iB^k)^2 +g \frac{b^2}{2a^2} + \frac{1}{\varphi}[(v^i)^2-a^2].
	\end{equation}
	
	Our construction provides the correct interaction between an external magnetic field and the spinning degrees of freedom. To verify this, we may look at the angular momentum $J_i$. The Lagrangian equations of motion imply
	\begin{equation}
		\dot{J}_i = \frac{e}{m}\ve_{ijk}J^jB^k,
	\end{equation}
	which is precisely the precession expected from a spinning particle under the effect of an exterior magnetic field.
	
	To conclude this Section, let us briefly comment on the quantization of the model presented here. The inner space of spin described above can be attached to space-time variables that describe the position of the spinning particle. It is done through the Lagrangian action
	\begin{equation}
		S = \int d\tau \left (\frac{m}{2}x^2 + e A_i \dot x^i - eA_0 + L'_{spin} \right ).
	\end{equation}
	In this case, $x^i$, $i=1,2,3$, are the space coordinates of the particle of mass $m$ and charge $e$. As usual, $B^i = \ve^{ijk} \partial _j A_k$; the second and third terms are just the minimal interaction with the external potential vector $A_i$, $A_0$. The transition to the Hamiltonian formulation leads to 
	\begin{equation}
		H = \frac{1}{2m}\left (p_i -e A_i \right )^2- \frac{e}{m}J_i B^i +eA_0.
	\end{equation}
	Hence, we quantize the model according to \eqref{quantization} and also act with $H$ on the state vector. We are left with 
	\begin{equation}
		i\hbar \frac{\partial \Psi}{\partial t} = \left [ \frac{1}{2m}\left (\hat{p}_i -e \hat{A}_i \right )^2- \frac{e\hbar}{2m}\sigma_i \hat{B}^i +e\hat{A}_0\right ] \Psi.
	\end{equation}
	As claimed, it is but the Pauli equation, where $\Psi$ stands for a two-dimensional spinor. All the details may be seen in \cite{Deriglazov_2010.sigma}. 
	
	\section{Local structures}\label{local.structure}
	
	Our examples are constructed with a peculiar freedom one has to write out Lagrangian models. Even when two Lagrangians differ, they may possess the same local structure, in the sense of being invariant under (almost) the same gauge transformations. Let us formalize these ideas that unite Lagrangians in equivalence classes. Our background shall be a configuration space parameterized by $\{ q^A\}$, $A\in \{1,...,N\}$. We write collectively
	\begin{equation}
		q^A = (x^a, y^\alpha,...).
	\end{equation}
	
	That is, the space can be split in different sectors. We may allow Lagrangians to depend only on a particular subset of variables, say, $L(x^a, \dot x^a)$. Let $\mathscr{L}$ be the set of all possible Lagrangians of the form $L = L(q^A, \dot q^A)$. Now, consider transformations $q^A(\tau) \rightarrow q'^A(\tau)$ parameterized by arbitrary functions $\epsilon^\alpha(\tau)$; $\alpha \in \{1,...,n\}$. Their infinitesimal form is given by
	\begin{equation}
		q^A \rightarrow q'^A = q^A + \sum^n_{k=0} \mathscr{R}^A{}_{(k) \alpha}(q^A, \dot q^A,...;\tau) \frac{d^k \epsilon^{\alpha}(\tau)}{d \tau^k},
	\end{equation}
	where the functions $\mathscr{R}^A{}_{(k) \alpha}$ are the corresponding generators. From here on, we write $\delta q^A:= q'^A - q^A$ for short.
	
	Let $L_1$ and $L_2$ be two Lagrangians in $\mathscr{L}$ invariant under gauge transformations $\delta_1 q^A$ and $\delta_2 q^A$, respectively. On $\mathscr{L} \times \mathscr{L}$, we set the following relation
	\begin{equation}\label{40}
		L_1 \sim L_2 \Leftrightarrow \delta_1 L_1 = \delta_2 L_2 + \frac{dF}{d \tau}
	\end{equation}
	whenever there exists such a function $F$. Clearly, $\sim$ is symmetric, reflexive and transitive, and as so, an equivalence relation. The corresponding equivalence classes will be designated by $[L] \in \mathscr{L}/\sim$ and for $L_1$, $L_2 \in [L]$, we say that the Lagrangians possess the same \textit{local structure}.
	
	Let us elucidate how this division of $\mathscr{L}$ in classes works with our previous examples.
	
	\textbf{Example 1.} Let us consider the already presented Lagrangians,
	\begin{align}
		L_1 &= \frac{m}{2}\eta_{AB}Dx^A D x^B,\\
		L_2 &= \frac{m}{2}\eta_{AB}Dx^A Dx^B - \xi g,
	\end{align}
	see Section \ref{introducing.local.invariance}.
	
	The two Lagrangians are invariant under the same transformations $\delta = \delta_1 = \delta_2$, given by
	\begin{equation}
		\begin{array}{cc}
			\delta x^A  = \frac{1}{2}\dot \alpha x^A - \alpha \dot x^A   \\
			\delta g = \frac{1}{2} \ddot \alpha -\dot \alpha g - \alpha \dot g
		\end{array}
	\end{equation}
	where $\alpha$ is an arbitrary function of the evolution parameter $\tau$. Clearly, $L_1 \neq L_2$, although a straightforward calculation leads to
	\begin{equation}
		\delta_2 L_2 = \delta_1 L_1+ \left ( \xi \alpha g - \frac{\xi \alpha}{2} \right )^\cdot \Rightarrow L_1 \sim L_2,
	\end{equation}
	that is, $L_1$ and $L_2$ bear the same local structure.
	
	\textbf{Example 2.} The second example is related to the spinning particle also presented in Sec. \ref{introducing.local.invariance}. In the first place, we take a Lagrangian defined on the configuration space $\{v^i, g \}$ given by
	\begin{equation}
		L_1 = \frac{1}{2g}(\dot v^i)^2
	\end{equation}
	while the second one is defined on an extended space $\{v^i, g, \varphi \}$ and is given by \eqref{lagrangian_spin}.
	In this case, the local symmetries are given by
	\begin{equation}
		\begin{array}{cc}
			\delta_1 v^i = \alpha \dot v^i, \\
			\delta_1 g = (\alpha g)\dot{},
		\end{array}
		\quad 
		\begin{array}{cc}
			\delta_2 v^i = \alpha \dot v^i,  \\
			\delta_2 g = (\alpha g)\dot{}, \\
			\delta_2 \varphi  = \alpha \dot \varphi - \dot \alpha \varphi.
		\end{array}
	\end{equation}
	$\alpha$ is an arbitrary function of the evolution parameter $\tau$. The local transformations lead to
	\begin{equation}
		\delta_2 L_2 = \delta_1 L_1 + \left ( \frac{\alpha g b^2}{2a^2}+ \frac{\alpha}{\varphi}[(v^i)^2 - a^2]\right )^\cdot \Rightarrow L_1 \sim L_2.
	\end{equation}
	This example shows an interesting feature of models with the same local structure. We can impose extra desirable constraints to a particular Lagrangian, enlarging the configuration space, maintaining the different Lagrangians so generated in the same class though. In this particular case, $L_1$ is not restricted to the sphere $(v^i)^2=a^2$ as is $L_2$. This constraint was introduced in $L_2$ by the variable $\varphi$.   
	
	\section{Conclusions}
	\label{sec.concl}
	
	This work discusses models with finite degrees of freedom that present local symmetries. Our new results are listed below. 
	
	\textbf{\textit{i.}} We have transposed the gauge principle from field theories to mechanical models. Although there are related works exhibited in the introduction, this is, up to our knowledge, the first attempt to do so with the generality hereby presented. In a few words, we start from a Lagrangian model invariant under a Lie group parameterized by constant factors. Then the group is raised to a local one and we require that the model stays invariant. With this prescription, the dynamical sector of the initial Lagrangian is now described by a covariant derivative, once a connection naturally appears throughout the method.
	
	\textbf{\textit{ii.}} Our construction was applied to different current models such as DSR proposals and classical spinning particles. Our prescription provides an original algebraic and geometric origin to the corresponding local invariance imposed, while giving insightful interpretations of relevant physical quantities that naturally emerge in the previously presented models. 
	
	\textbf{\textit{iii.}} Finally, and still concerning the local symmetries, we have shown how the set of Lagrangians is partitioned in equivalence classes. The corresponding equivalence relation was defined in \eqref{40}. We say that two Lagrangians possess the same local structure when they differ by a total derivative term under local transformations. This composition of the set $\mathscr{L}$ of Lagrangians was particularly important to our own examples. In the first case (regarding DSR proposals), we could deform the initial Lagrangian to a new one, preserving the local structure. In the new framework, the invariant scale $\xi$ presented in DSR models appears as a coupling constant to the gauge variable. The second example (concerning spinning particles) shows how we can add extra variables to the configuration space, enlarging the initial configuration space while still conserving the local structure. For that case, the last term added guaranteed a desirable constraint.  
	
	Furthermore, this work may shine some light on the importance of formalizing key aspects of classical mechanics, such as the gauge principle for models with finite degrees of freedom. Even though a more complete description of mechanics is accomplished with quantum field theory, approaches like the one presented here play a central role in the finding and understanding of the physical sector of important models formulated with extra degrees of freedom, thus justifying their formalization.

	\section*{Acknowledgments}
	
	BFR would like to thank G. Caetano for indicating valuable references concerning the analogy between non-Abelian gauge theories and general relativity. 
	
	This work was supported by grant \#2021/09311-5, São Paulo Research Foundation (FAPESP) and Programa Institucional de Bolsas de Iniciação Cientí-fica - XXIX PIBIC/CNPq/UFJF - 2020/2021, project number ID47862.
	
	GFVJr was a CNPq - Brazil fellow during the first months of production of this work.

\end{document}